# Quasi-three-level laser emissions of neodymium doped disordered crystal waveguides

Yang Tan, Feng Chen, Javier Rodríguez Vázquez de Aldana, Haohai Yu and Huaijin Zhang

*Abstract*—This paper reports on the quasi-three-level continuous wave laser operation based on waveguide structures in neodymium doped calcium niobium gallium garnet disordered crystal. Laser wavelength selection through the waveguide cross section was observed. Waveguide structures with different cross sections were fabricated by the ultrafast laser inscription, which have propagation losses around 1 dB/cm. With suitable pumping conditions, laser emissions were observed at the low-wavelengths of ~930 nm and ~890 nm. The lasing threshold for the low-wavelength emission was around 50 mW, which is far below the threshold of several watts reported in the bulk laser system. In addition, it was found the laser generation at the wavelength of ~890 nm has direct relationship with the volume of the waveguide structure. The results suggest advantages of the waveguide platforms over the bulk systems on the low-wavelength laser emission.

AMONG the family of neodymium-doped disordered laser materials, neodymium ion doped calcium niobium gallium garnet (Nd:Ca$_3$(NbGa)$_{2-x}$Ga$_3$O$_{12}$ or Nd:CNGG) is an attractive gain medium. Owing to the random distribution of the niobium and gallium ions, it has lower melting point of 1460 ℃, broader absorption band and larger emission bandwidth compared with ordered crystals, which makes it suitable for multi-wavelength and ultra-short laser [1-3]. Based on the Nd:CNGG bulks, laser emissions at the wavelength of ~ 930 nm [4], 1.06 μm [5] and 1.33 μm [6] have been realized.

For the Nd-doped medium, the laser emission with wavelength less than 950 nm (so-called low-wavelength) could be generated through the quasi-three-level laser transition process. Low-wavelength lasers can be used for the blue spectrum laser emission combined with the second-harmonic conversion process [7]. For example, the fundamental wavelength of 800 nm~950 nm could be transformed into the spectrum of 400 nm~445 nm. Compared with semiconductor lasers, it has smaller spectral width and better coherence length, which makes it interest for interferometer experiments. Besides, the quasi-three-level laser transition around 900 nm has the potential application as the sources for the differential absorption lidar (DIAL) in water vapor detection, which have wavelengths overlap the absorption peaks of water vapor. [8] As reported in Ref. [4, 9, 10], Nd:CNGG disordered crystal has proved to be an efficient gain medium for the laser oscillation around 930 nm. However, No observation on the laser emissions at ~ 890 nm has been reported in Nd:CNGG crystal as of yet.

Waveguide lasers, taking advantages of the compact platform of waveguide structures, are promising integrated light sources [11-16]. With appropriate design of dielectric mirrors, lower lasing thresholds and considerable pumping efficiencies can be easily obtained compared with the bulk lasers. The waveguide structure in Nd:CNGG was firstly fabricated by ion implantation, but the propagation loss was high (3 dB/cm), which limited the laser emissions from the structure [17, 18].

The femtosecond (fs) laser inscription is a robust, mature technique for waveguide fabrication in a number of gain media, including glasses, ceramics and single crystals [19-26]. In crystalline materials, there are three configurations of fs laser inscribed waveguides, i.e., Type I waveguides (with positive refractive index changes in the track), Type II stress-induced waveguides (typically located in the region between two tracks with reduced indices) and Type III depressed cladding waveguides (surrounded by low-index tracks). Compared with Type I waveguides of crystals, the Type II waveguides are more stable. In addition, the modal feature of Type II was detected to be a single mode and has smaller size compare with Type III waveguide. Hence, Type II structures are usually superior to those of cladding waveguides. So far, the highest output power of waveguide laser was obtained based on Type II structure [23]. According to our previous work [27], four-level Nd-ion transition from Type III waveguide was observed. But no quasi-three-level laser operation has been observed in the waveguide structure.

In this Letter, we reported on the fabrication of fs laser inscribed Nd:CNGG Type II waveguides. Properties of waveguide, such as modal profiles and propagation losses, were obtained. Under certain pumping conditions, the stable laser oscillations at low-wavelengths of ~ 930 nm and ~ 890 nm

This work is carried out under the support by the National Natural Science Foundation of China (No. 11274203), the Spanish Ministerio de Ciencia e Innovación (Projects CSD2007-00013 and FIS2009-09522), and Junta de Castilla y León (Project SA086A12-2). Tan acknowledges the support by the Independent Innovation Foundation of Shandong University (IIFSDU, No.104222012GN056 / 11160072614098) and China Postdoctoral Science Foundation (Grant No. 2013M530316).

Yang Tan and Feng Chen are with the School of Physics, State Key Laboratory of Crystal Materials and Key Laboratory of Particle Physics and Particle Irradiation (Ministry of Education), Shandong University, Jinan 250100, China (e-mail: tanyang@sdu.edu.cn and drfchen@sdu.edu.cn).

Javier Rodríguez Vázquez de Aldana is with Departamento Física Aplicada, Facultad Ciencias, Universidad de Salamanca, Salamanca 37008, Spain.

Haohai Yu and Huajin Zhang are with the State Key Laboratory of Crystal Materials, Shandong University, Jinan 250100, China.

were realized inside the waveguides.

## I. EXPERIMENTS OF WAVEGUIDE FABRICATION AND LASER EXCITATION

The Nd:CNGG crystal used in this work is the same as the one reported in Ref. [9], which was doped by 2 at.% $Nd^{3+}$ ions and cut into dimensions of 1.5 ×8 ×9.9 ($x \times y \times z$) mm$^3$ with all the facets optically polished. A linearly polarized 120 fs pulsed laser (Spitfire, Spectra Physics, USA) was used for the inscription of damaged tracks, with a central wavelength of 800 nm, a repetition rate of 1 kHz and a maximum pulse energy of 1 mJ. During the inscription process, the energy of the pulses was reduced to 0.84 μJ and was focused inside the sample with a 40 × microscope objective, scanning the sample at 25 μm/s. Two parallel tracks were thus written with different separations. As depicted in Figs. 1(a) and 2(b), waveguide 1 and waveguide 2 (W1 and W2) were formed between the tracks with width of 20 μm and 15 μm, respectively.

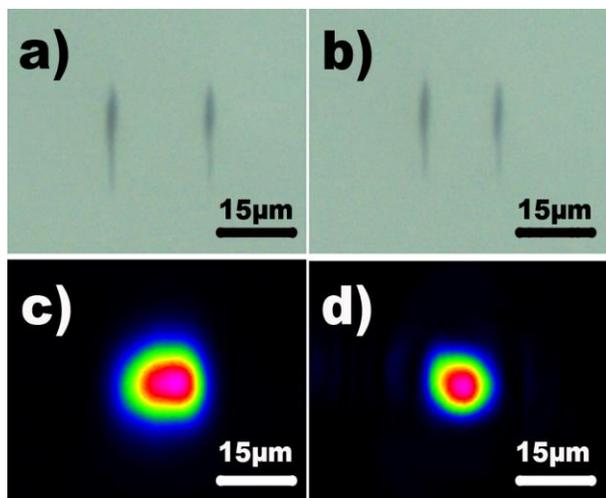

Fig. 1. Cross section of waveguides (W1 and W2) with widths of 20 μm (a) and 15 μm (b); the measured intensity distribution of the $TM_{00}$ modes at the wavelength of ~930nm (c) and ~890 nm (d).

Figures 1 (c) and 1(d) show the modal profiles of the waveguide at the wavelengths of ~ 930 nm and ~ 890 nm, respectively, showing single-mode waveguiding features. Adjusting the polarization of the detecting light, we found only TM (parallel to tracks) polarized light could be confined inside the waveguide. To avoid the absorption from the doped rare-earth ions, a laser at wavelength of 632.8 nm was used to measure the propagation loss by the method described in Ref. [28]. The value of propagation loss was determined to be less than 1 dB/cm.

We measured the room-temperature luminescence spectra obtained from the W1, W2 and bulk. During the measurement, pumping laser at 810 nm was coupled into waveguides and the bulk, respectively. Meanwhile, the output light was collected and detected by an optical spectrograph (measurement error less than 2 nm). As shown in Fig. 2, shapes and position of fluorescence peaks were almost identical, which means the fs laser inscription process did not affect the fluorescence properties of the Nd:CNGG crystal.

Figure 3 demonstrates the experimental setup for the laser excitation. Through a convex lens (focal length 20 mm), a continuous laser at the wavelength of 810 nm from a wave tunable Ti:Sapphire laser was coupled into the Nd:CNGG waveguide. To build the resonant optical cavity for the laser oscillation at the low-wavelength, specially designed Mirror 1 and Mirror 2 (M1 and M2) were adhered to the end-facets of the waveguide.

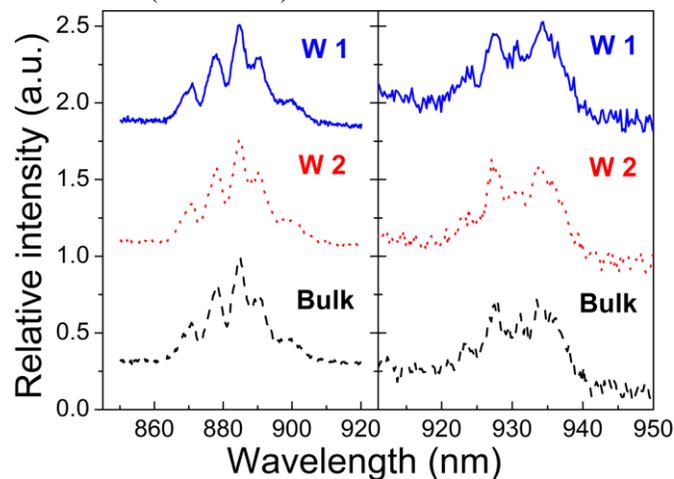

Fig. 2. Luminescence emission spectra of $Nd^{3+}$ ions at $^4F_{3/2} \rightarrow {^4I_{9/2}}$ transition obtained from W1 (blue solid line), W2 (red dotted line), and the bulk (black dashed line).

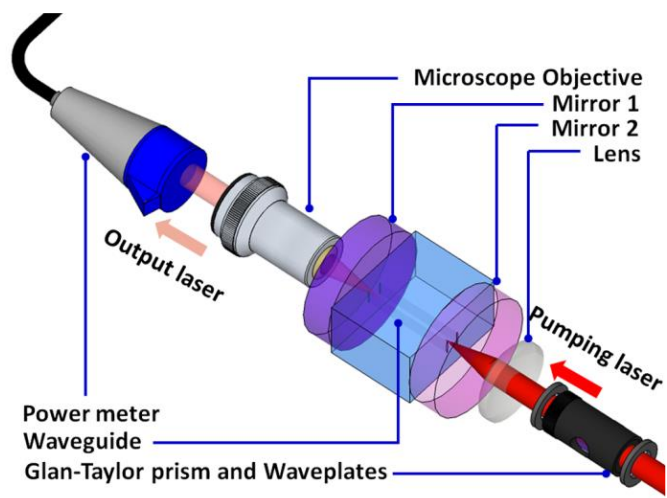

Fig. 3. Experimental setup for the waveguide laser excitation. Mirror 1 (M1) and Mirror 2 (M2) were spectral designed for certain wavelength laser excitation.

Figure 4 depicts the schematic of the pumping together with upper lasing levels of the $Nd^{3+}$ ions. In the Nd-doped medium, the lasing at ~930 nm comes from the quasi-three-level transition, which corresponds to the emission of $^4F_{3/2} \rightarrow {^4I_{9/2}}$. As one can see, there is the same upper laser state ($^4F_{3/2}$) for the ~ 890 nm and the ~ 930 nm transition, which indicates that the laser emission at the wavelengths of ~ 890 nm and ~ 930 nm could be excited respectively. In order to select the laser emission at the certain wavelength, the reflectivity of the mirrors was accurately selected. For example, M1 (M2) has 90% (99%) reflectivity at the wavelength of ~930 nm to realize the laser emission at that wavelength. In addition, a pair of mirrors with 90% (M1) and 99% (M2) reflectivity at ~ 890 nm was used for the ~ 890 nm wavelength laser oscillation.

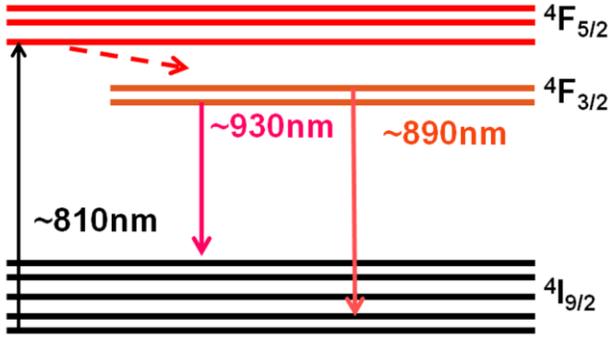

Fig. 4. Schematic description of the energy structure of Nd:CNGG crystal.

## II. RESULTS AND DISCUSSION

Using the pair of ~ 930 nm mirrors, we obtained the laser oscillation from W1. The inset of Fig. 5 depicts the spectrum of the output laser from W1, when the pumping laser power was above the laser threshold. As one can see, there is a dual-wavelength laser oscillation at the wavelengths of 927.5 nm and 934.1 nm.

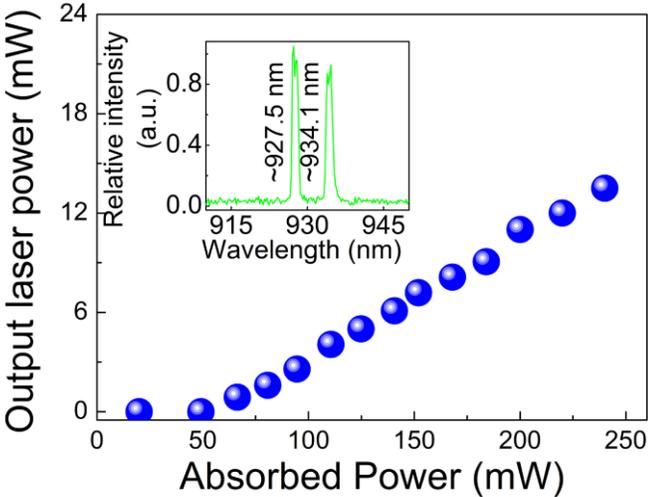

Fig. 5. Power of the total output waveguide laser versus the absorbed pumping power. The inset shows the cw laser oscillation spectra from W1 after pumping at 810 nm above the power threshold.

Pumping laser with different polarization was used to pumping waveguide laser in this work. With the TM polarization of pumping laser, the laser threshold was around 50 mW and the maximum output power was around 14 mW with ~ 240 mW absorbed pumping laser. The slope efficiency was around 7.4 %. However, the slope efficiency of the waveguide laser was just 2% and the lasing threshold was up to 100mW with the TE polarized pump laser. This may be because the Type II waveguide is induced by the compression form tracks, which could lead to the birefringence phenomenon in the waveguide region. And light with TM polarization will has higher coupling efficiency compared with TE light. Hence, more pumping light with TM polarization could be coupled into waveguide with the same coupling power.

Compared with the bulk Nd:CNGG laser, an additional wavelength at ~ 927.5 nm was found and the threshold was far below the value (3 W) in the bulk [9]. We believe the enhanced laser operation is induced by the waveguide structure, which has a much smaller volume of the resonant cavity.

The mirrors were then replaced by the pair for ~890 nm. From W2, ~ 890 nm laser emission was obtained when the power of absorbed laser was more than 70 mW with TM pumping laser. As depicted in Fig. 6, a dual-wavelength laser oscillation was obtained at the wavelengths of 885 nm and 891 nm, respectively. The threshold for both wavelengths was similar, and the maximum output power was around 10 mW when the pumping power was ~ 240 mW at the wavelength 810 nm with TM polarization. This is the first time laser oscillation at ~ 890 nm was observed based on Nd:CNGG crystal.

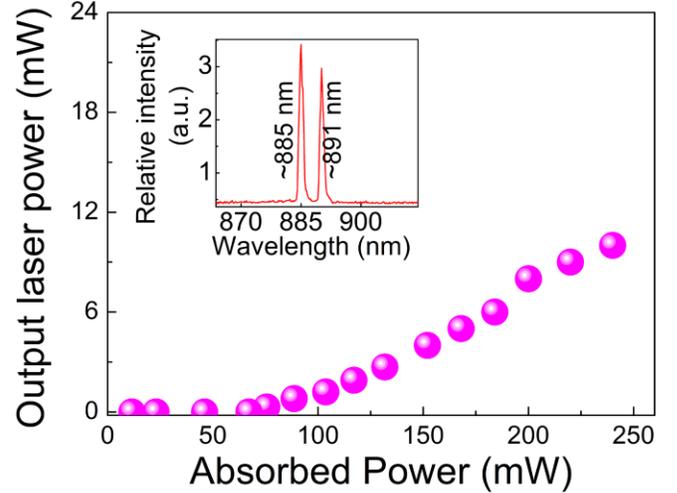

Fig. 6. Output laser power as a function of the absorbed pumping power. The inset is the cw laser oscillation spectra from W2 with pumping at 810 nm above the pumping power threshold.

Surprisingly no laser emission at ~ 890 nm was found in W1 with the same mirrors and pumping conditions. Similar thing was also observed in Nd:CNGG Type III waveguide [27]. With the same mirrors and pumping conditions, no quasi-three-level laser could be excited. The main difference of these waveguide is the size of volume of the waveguide. It seems that the volume of the waveguide structure directly affects the laser performance for the quasi-three-level wavelength.

To explain this phenomenon, we propose to use the same model as for the three-level laser in Yb-doped materials [29]. In order to achieve laser emission, the gain coefficient should be considered as the main quantity, which is related to the effective cross sections at the laser wavelength ($\lambda$) including the absorption cross section ($\sigma_{abs}(\lambda)$) and the emission cross section ($\sigma_{em}(\lambda)$). As ground state resonance absorption is present at the laser wavelength for Nd lasers, the minimum fraction ($\beta_{min,\lambda}$) of excited Nd ions must balance the ground-state absorption and the gain exactly. For the lowest wavelength, $\beta_{min,\lambda}$ could be expressed by equation [29]:

$$\beta_{min,\lambda} = \frac{\sigma_{abs}(\lambda)}{\sigma_{abs}(\lambda) + \sigma_{em}(\lambda)} \qquad (1)$$

To achieve transparency at the extraction wavelength, we could calculate the minimum required pump power ($P_{min}$) by the equations:

$$I_{sat} = h\upsilon /(\sigma_{pump,abs}\tau_{em}) \quad (3)$$

$$P_{min,\lambda} = \beta_{min,\lambda} I_{sat} S \quad (4)$$

where $I_{sat}$ is the pump saturation intensity parameter; $\sigma_{pump,abs}$ and $\tau_{em}$ are the absorption cross section and the emission life-time at the pumping wavelength; and $S$ is the waveguide cross section.

For the situation of the quasi-three-level laser transition in Nd:CNGG crystal, $\sigma_{em}(890)$ and $\sigma_{abs}(890)$ are $1.8 \times 10^{-20}$ cm$^2$ and $0.83 \times 10^{-20}$ cm$^2$ [5] respectively. According to Eq. 1, $\beta_{min,880}$ is around 0.68. $\sigma_{abs}(930)$ is supposed to be less than $0.2 \times 10^{-20}$ cm$^2$, while $\sigma_{em}(930)$ is about $0.8 \times 10^{-20}$ cm$^2$. Hence, $\beta_{min,930}$ should be less than 0.18.

As shown in Eq. (3), the pumping threshold power ($P_{min,\lambda}$) is proportional to $\beta_{min,\lambda}$ and $S$. Based on the above calculation, the pumping threshold at ~ 890 nm should be more than three times larger for laser emission at 930 nm in the waveguide with the same cross section ($S$). Therefore an efficient way to decrease the pumping threshold is to decrease the area of the cross section for the laser emission with higher value of $\beta_{min,\lambda}$.

The stability of the waveguide laser was discussed. We measured the variation of output laser power by a photodiode detector and an oscilloscope with a sampling interval of 2 ns. The variation ratio of the output light was less than 0.2% measured over a period of 10 min, which means the Nd:CNGG waveguide laser is a stable continuous-wave laser source.

In conclusion, low-wavelength laser was generated based on Nd:CNGG waveguides, which was fabricated by the ultrafast laser inscription method. The propagation in the waveguides was single-mode and the measured loss was 1 dB/cm. The power thresholds of the waveguide lasers was ~ 50 mW and ~ 70 mW for the wavelengths of ~ 930 nm and ~ 890 nm respectively. The stability of this waveguide laser was discussed and it was proved to be a continuous stable laser. Nevertheless, future effort would be performed on the improvement of the efficiency of the Nd:CNGG waveguide system.

**Yang Tan** is currently a Lecturer of School of Physics, Shandong University, China. He received the Ph.D. degree Shandong University in 2011. His research interests include material modifications by ultrafast lasers and ion beam, nonlinear optics, optical waveguides, etc.

**Feng Chen** is currently a Professor and the Head of School of Physics, Shandong University, China. He received the Ph.D. degree Shandong University in 2002. He was with Clausthal University of Technology, Germany, from 2003 to 2005, as an Alexander von Humboldt Research Fellow. He became a Professor at Shandong University in 2006. His research interests include material modifications by ultrafast lasers and ion beams, optical waveguides, etc. Prof. Chen is a Fellow of Institute of Physics (IOP), UK, a Senior Member of the Optical Society of America (OSA) and Chinese Optical Society (COS). He also serves as an Associate Editor of *Optical Engineering*.

**Javier R. Vàzquez de Aldana** received the degrees of Bachelor of Science (1997) and Ph.D. (2001) from the University of Salamanca, Spain. He is currently Associate Professor of the Science Faculty, University of Salamanca, Spain. His research activity is focused on the interaction of intense femtosecond pulses with materials and its application to the fabrication of photonic devices. He is a member of the Laser Microprocessing Research Group, and is also technical and scientific advisor of the Laser Facility at the University of Salamanca.

**Haohai Yu** was born in Jinan, China, on October 16, 1981. He received the Ph.D. degree from Shandong University, Jinan, in 2008. He is currently an Associate Professor at State Key Laboratory of Crystal Materials and Institute of Crystal Materials, Shandong University. His current research interests include crystal growth, diode-pumped solid-state lasers, and nonlinear optics based on the new crystals.

**Huaijin Zhang** was born in Shandong, China, in 1965. He received the B.S., M.S., and Ph.D. degrees in physics from Shandong University, Jinan, China, in 1988, 1994, and 2001, respectively. He is currently a Professor of the State Key Laboratory of Crystal Materials, Shandong University. His current research interests include the growth and characterization of different types of crystal materials, including laser, nonlinear, electrooptic, and piezoelectric crystals.